\documentstyle[sprocl,psfig]{article}

\begin{document}

\title{Atmospheric Neutrinos: Past, Present and ...}

\author{John M.~LoSecco}

\address{Physics Department, University of Notre Dame\\
Notre Dame, Indiana 46556-5670 USA\\
E-mail: losecco@nd.edu}

\maketitle\abstracts{This paper reviews the history of the atmospheric
neutrino anomaly.  The historical record does not support more recent claims
that the anomaly constitutes evidence for a neutrino mass.  Most experiments
have reported an apparent muon deficiency which is independent of energy
and distance.  Time dependent variations in the Super Kamioka results are
noted.}
\section{Prehistory}\label{sec:preh}
\subsection{Opportunity}\label{subsec:opp}
The early history of the atmospheric neutrino anomaly is closely tied to the
growth of grand unified theories.  Grand unified theories inspired the
experimental search for proton decay which led to large well shielded
detectors.  The primary background to proton decay was due to atmospheric
neutrinos so they were studied in some detail.  Neutrino oscillations figured
prominently at the first workshop on grand unification.  About half of the IMB
paper at that meeting~\cite{sulak} was devoted to the question of studying
neutrino oscillations with atmospheric neutrinos.
\subsection{Early Indications}\label{subsec:early}
Early indications that atmospheric neutrinos were not behaving as expected
came because many interesting proton decay modes produced unstable particles
such as kaons and muons that could be expected to decay in the detector and
yield a delayed coincidence with the primary event.  In looking for these
delayed coincidences it was noticed that there appeared to be too few of
them in the global data sample.  Based on 148 events Shumard
reported~\cite{shumard} that 26$\pm$4\% of the observed events had a muon
decay.  Careful studies had indicated that a decay rate of 35$\pm$1\% was
expected.  This 2.1 sigma difference was discounted in the thesis.

Confirmation came from the Kamioka experiment in Japan.
Kajita's thesis~\cite{kajita} had evidence for a 2.4 sigma deficit of muon
decays based on 89 single ring events,  But no note of the deficit was made
in the thesis.

\begin{figure}[t]
\centerline{\mbox{\psfig{figure=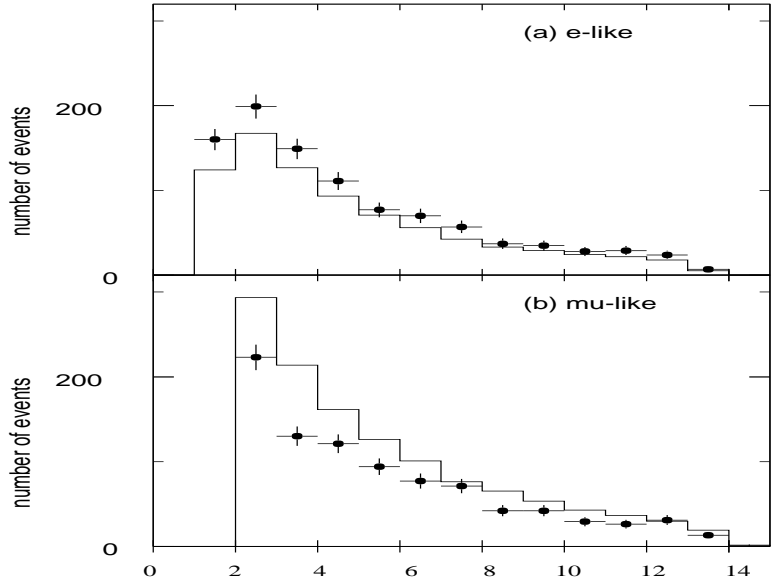,height=3in,width=4in}}}
\caption{The electron and muon neutrino spectra as reported by the
Super Kamioka experiment\label{fig:spec}}
\end{figure}
The atmospheric neutrino statistics rapidly increased and the deficit was
noted in a 1986 IMB publication~\cite{haines}.  A 3.3 sigma difference was
reported based on 401 events.  26$\pm$2\% of the events contained a muon
decay when 34$\pm$1\% was expected.
\subsection{Flux Modeling}\label{subsec:model}
In all of these cases there were suspicions about the {\em expected}
values.  The atmospheric neutrino flux, the neutrino interaction model and
the detector response were all required for an estimate.  Substantial effort
went into modeling the neutrino interactions themselves.  (This is because
the high multiplicity neutrino interactions were the most serious background
to the proton decay signal).
Shumard had used Freon bubble chamber data as a model of neutrino
interactions. Kajita and Haines had used neutrino
interaction models which they compared with deuterium bubble chamber data.

Except for the muon decay rate deficiency the observations were all in good
agreement with expectations of atmospheric neutrinos.
The rate of interactions, their homogeneity in the detector and the
{\em isotropy} of the signal were as expected~\cite{imb1,kamnu}.
\section{The Classical Period}\label{sec:class}
\subsection{Rapid Progress}\label{subsec:rapid}
The Kamioka group applied a {\em revised} form of their particle identification
methods to follow up on these indications to establish the general
properties of the anomaly~\cite{kamd}.  The early work implied that the effect
could be due to either a deficit of muon neutrinos or an excess of electron
neutrinos.  Uncertainties in the flux normalization did not permit one to
discriminate between these two possibilities.  To reduce the dependence on
the absolute flux normalization the Kamioka group introduced a variable
called $R$, $R= (\frac{\mu}{e})_{Obs} / (\frac{\mu}{e})_{Expect}$, which was
only sensitive to the relative normalization of the muon and electron neutrino
flux.

Typical of the results from this period is an apparent energy independent
excess of electron type events and an apparent energy independent deficit of
muon type events.  This is illustrated in figure \ref{fig:spec} which is taken
from the
Super Kamioka early work~\cite{feb98}.  The solid curve is the expected value
the points are the data.  This confirms prior
work~\cite{kamd,casper,kam92} on these properties.
\begin{figure}[t]
\centerline{\mbox{\psfig{figure=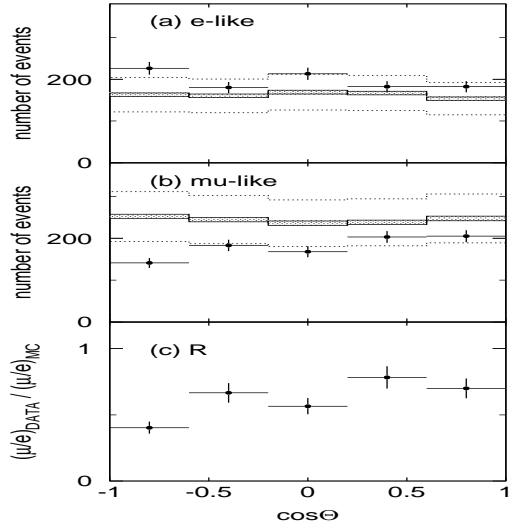,height=3in,width=3in}}}
\caption{The electron and muon neutrino zenith angle distribution as reported
by the Super Kamioka experiment.  The bottom plot is the zenith angle
distribution for $R$.\label{fig:zenith}}
\end{figure}
These data have $R=0.61\pm0.03(stat.)\pm0.05(sys.)$ which is well below the
expected value of 1.  This value of $R$ seems to be energy
independent~\cite{feb98} over the range of energies illustrated.

The angular distribution of the effect is very important in testing neutrino
evolution effects.  Atmospheric neutrinos from above have traveled a few 10's
of km.  Those from below have traveled on the order of 10,000 km.
$R$ as a function of zenith angle from early Super Kamioka work~\cite{feb98}
illustrates the most important features of the angular distribution.

While IMB~\cite{casper} and Kamioka~\cite{kamd,kam92} have shown purely
isotropic distributions the most important feature is still present in 
figure \ref{fig:zenith}, the Super Kamioka plot.  Note that the electrons
are high and the muons are low at all angles.
The value of $R$ is significantly low over all of the
solid angle, even at $0^{\circ}$.  This implies that even atmospheric
neutrinos with
short path lengths, on the order of 10's of kilometers, are effected by the
anomaly.
\subsection{Uniformity in Energy and Direction}\label{subsec:details}
The isotropic angular distribution could be misleading.  The direction
reconstructed and plotted is the charged lepton direction emerging from the
neutrino interaction.  This has a reasonable correlation with the incident
neutrino direction, but it is not exact~\cite{sulak}.  Two factors mitigate
the significance of this scattering
effect.  The directional correlation improves with energy and the
distance scales very slowly with angle except near the horizon.  The majority
of the atmospheric neutrino events from above have traveled on the order of
10's of km.  Those from below have traveled the order of 10,000 km.  So
scattering within the hemisphere has only a small effect on the distance
traveled.  With regard to the first factor, the data indicates~\cite{takita}
that the isotropy for both the electron and muon samples is still present at
higher energies.  See figure \ref{fig:updn}.

\begin{figure}[t]
\centerline{\mbox{\psfig{figure=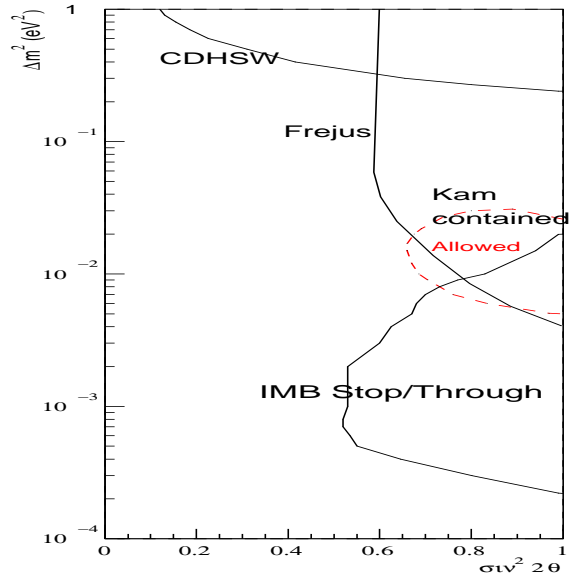,height=3in,width=3in}}}
\caption{The exclusion plot for $\nu_{\mu} \rightarrow \nu_{\tau}$
from a number of experiments.  The IMB Stop/Through excludes the region
favored by recent Super Kamioka analyses and much of the region permitted
by the older Kamioka analysis too.\label{fig:ecl}}
\end{figure}

Since neutrino oscillations were always a potential hypothesis for such an
effect and since there was no apparent energy or distance dependence in the
contained neutrino events another source of signal could be used to extend
the sensitivity or corroborate the oscillation hypothesis.
Figure \ref{fig:ecl} shows the
exclusion plot for the $\nu_{\mu} \rightarrow \nu_{\tau}$ hypothesis using
upward going neutrinos~\cite{upward} and the ratio of stopping to through-going
upward muons~\cite{svob}.  The Frejus limits~\cite{frejus} are also shown.
Most large mixing angle $\Delta m^{2}$ were ruled
out by the combination of analyses reviewed in this paper~\cite{svob}

\subsection{Consistency}\label{subsec:const}
By early 1998 the anomaly had been consistently observed in about 8
independent measurements.  All the observations were consistent with an
$R$ value of 0.61.

\begin{tabular}{lc} 
Experiment & Measured R value\\
\hline
Kamiokande  Sub-GeV & 0.60$\pm$0.07$\pm$0.05\\
Kamiokande  Multi-GeV& 0.57$\pm$0.08$\pm$0.07\\
IMB & 0.54$\pm$0.05$\pm$0.12\\
Frejus & 1.00$\pm$0.15$\pm$0.08\\
Nusex & 0.99$\pm$0.29\\
Soudan & 0.64$\pm$0.17$\pm$0.09\\
Super Kamiokande  Sub-GeV & 0.61$\pm$0.03$\pm$0.05\\
Super Kamiokande  Multi-GeV& 0.66$\pm$0.06$\pm$0.08\\
\end{tabular}
\subsection{Up/Down Energy Independence}\label{subsec:indep}
The atmospheric neutrino flux is not truly isotropic.  The Earth's magnetic
field, which varies from point to point limits the flux of primary cosmic
rays which can get close enough to the atmosphere to interact.  Fortunately
at some sites, such as the IMB site, the effect is small.  IMB was located
at 81.27$^{\circ}$ west and 41.72$^{\circ}$ north
where the the downward flux was expected to be about 5\% greater than that
coming from below.  The geomagnetic effects decrease with energy (but are
replaced by path length effects at high energy.)  In setting limits at IMB
it has been customary to neglect the small geomagnetic effects.  Any reduction
in the upward flux due to geomagnetic effects would be attributed to the
phenomena in question.  This neglect of geomagnetic effects at IMB makes the
limits obtained slightly conservative.  In other words in setting limits for
neutrino oscillations a larger region could have been excluded if geomagnetic
effects were taken into account.

\begin{figure}[t]
\centerline{\mbox{\psfig{figure=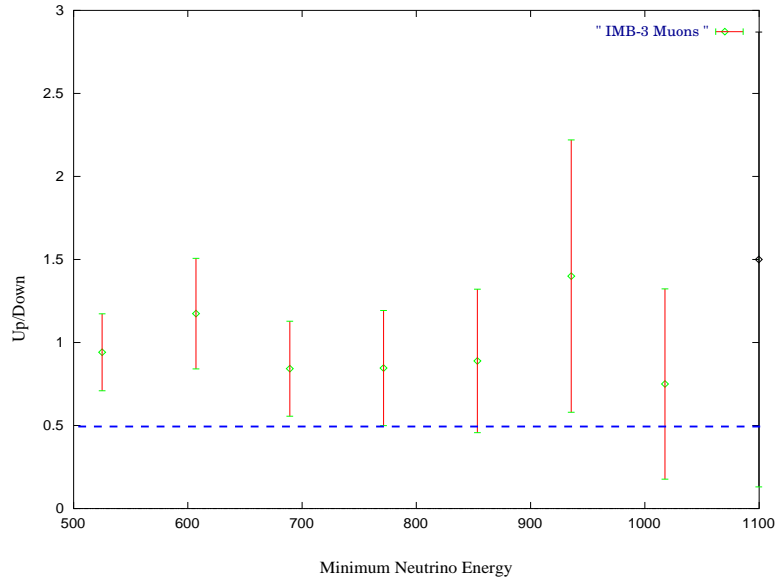,angle=270,height=3in,width=4in}}}
\caption{The upward over downward muon rate in IMB3 plotted as a function
of the minimum energy.  There is no indication of an asymmetry at any
energy.\label{fig:updn}}
\end{figure}

Limits based on the up/down rate for muon neutrino interactions in
IMB-3~\cite{casper} are shown in figure~\ref{fig:rat}.  The up over down ratio
is shown as a function of the minimum neutrino energy in figure~\ref{fig:updn}.
While the statistical errors grow with energy as fewer events are used, the
data are consistent with an energy independent value of 1.  There is no
evidence for an up-down asymmetry in this data.
\newsavebox{\figecl}
\sbox{\figecl}{\ref{fig:ecl}}
\begin{figure}[t]
\centerline{\mbox{\psfig{figure=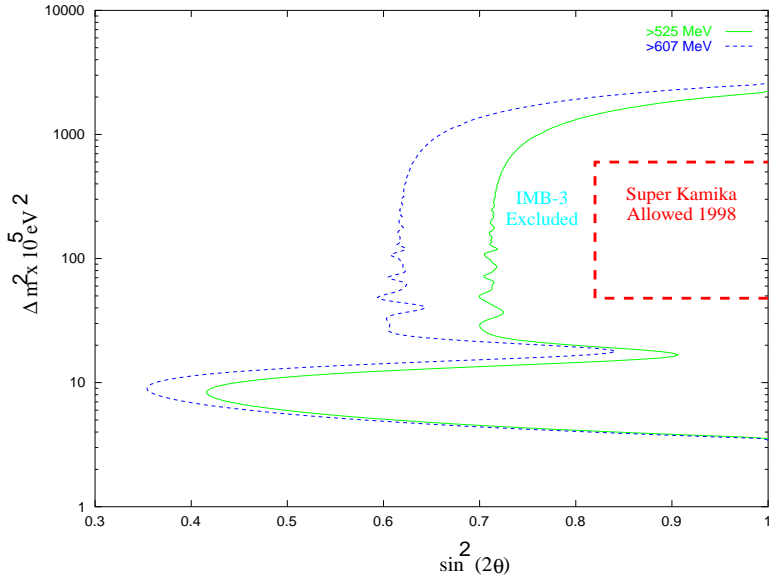,angle=270,height=3.in,width=4in}}}
\caption{The excluded region extracted from the IMB3 upward
over downward
muon rate.  The region favored by the Super Kamioka experiment is ruled out
by this measurement and the one illustrated in figure~\usebox{\figecl}.
The larger excluded region uses all of the muon neutrino events.  The
smaller region uses those events above about 600 MeV.\label{fig:rat}}
\end{figure}
\section{The Modern Era}
\begin{figure}[t]
\centerline{\mbox{\psfig{figure=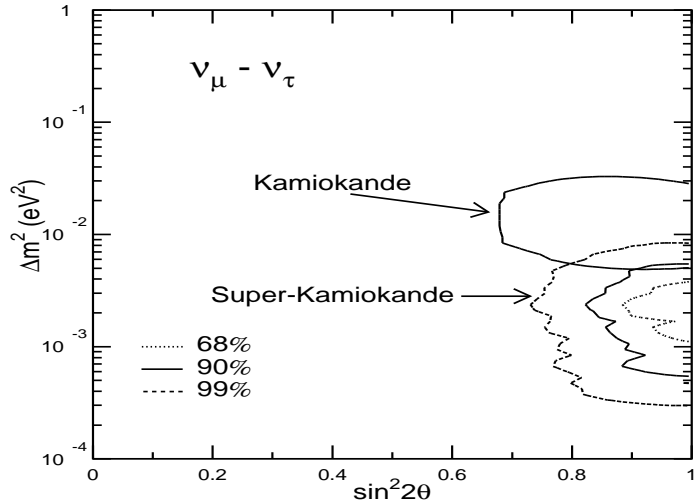,height=3in,width=4in}}}
\caption{The Super Kamiokande and Kamiokande allowed regions for
$\nu_{\mu} \rightarrow \nu_{\tau}$ oscillations.\label{fig:alow}}
\end{figure}

The hint of a directional modulation, present in figure \ref{fig:zenith}
apparently evolved and in mid 1998 the Super Kamioka collaboration
published~\cite{nuosc} it as evidence to support the neutrino oscillation
hypothesis (figure~\ref{fig:alow}).
This result has been widely discussed and interpreted.
It is noteworthy that the mass fit to the Super Kamioka
data was barely consistent with the earlier Kamioka neutrino analysis.
The reason for this difference is that the two data samples are indeed
different.  Earlier experiments, including Kamioka do not corroborate
the Super Kamioka observations of anisotropy.  The anomaly is still manifest
at 0$^{\circ}$ in this data~\cite{nuosc}.  It can not be accommodated by the
$\Delta m^{2}$ of the fit but is handled by several other fit parameters.
Both the $\mu$ to $e$ flux ratio and the up to down flux ratio are fit as
part of the oscillation analysis.
\section{The Post Modern Era}
The modern period did not last long.  In January 1999 the Super Kamioka
group presented~\cite{dpf}  additional data to increase the exposure from
414.4 days to 736 days.  The data was presented in a manner to bolster
confidence in the neutrino oscillation hypothesis.  But under closer
examination~\cite{vary} the additional data appeared to be {\em inconsistent}
with the sample that had already been published.

Without detailed access to the data it is hard to understand the exact
nature of the change.  Manifestations of the problem are a 12$\pm$3\% drop
in overall event rate for the sub-GeV event sample.  Most of the drop
seems to be due to a 18$\pm$5\% drop in the ``electron neutrino'' interaction
rate and a 20$\pm$5\% drop in the multi-ring rate.  The muons manifest a not
significant 3$\pm$5\% rise.  There is no significant change in the
multi-GeV data sample.

Since the observed ``electron neutrino'' rate has dropped there is a
significant
change in $R$ over values reported earlier.  The value of $R$ reported in
the new sub-GeV sample was, $0.76\pm0.04(stat.)\pm0.06(sys.)$.  The
earlier value was $R=0.61\pm0.03(stat.)\pm0.05(sys.)$.  Only the statistical
error is relevant in comparing these two numbers since they are taken from the
same experiment and share the same systematic error.  To the extent that the
atmospheric anomaly is $R \ne 1$ this variation is rather shocking.
\section{What Next?}
The apparent temporal variation may be a significant factor in understanding
what is actually happening with the anomaly.  We have been assured that the
temporal effect is not
a result of systematic error.  If it were attributable to systematic error
it would invalidate much of the Super Kamioka physics but not help in
furthering our understanding of the effect.
The Super Kamioka group has shown evidence~\cite{kasuga} of the stability of
their detector.  Previous experiments such as Kamioka~\cite{takita} and IMB
have found no indications of temporal variation.  There is some
evidence~\cite{vary} that the change in $R$ is continuing and has not yet
reestablished a constant value.
\section*{Acknowledgments}
I am grateful to S.~and G.~Domokos for their hospitality at this meeting.
I would like to thank I.~Bigi for encouragement.


\section*{References}
\begin{thebibliography}{99}
\bibitem{sulak}L.~Sulak in {\em First Workshop on Grand Unification},
ed. P.~Frampton, S.~Glashow and A.~Yildiz (Math Sci Press, Brookline, Mass.
1980).
\bibitem{shumard}E.~Shumard, Ph.D. thesis, University of Michigan (1984).
\bibitem{kajita}T.~Kajita, Ph.D. thesis, University of Tokyo (1986).
\bibitem{haines}T.~Haines {\em et al.} Phys.~Rev.~Lett.~{\bf 57} 1986 (1986).
\bibitem{imb1}R.M.~Bionta {\em et al.} Phys.~Rev.~{\bf D38}, 768 (1988).
\bibitem{kamnu}M.~Nakahata {\em et al.} J.~Phys.~Soc.~Jap.~{\bf 55} 3786
(1986).
\bibitem{kamd}K.~Hirata {\em et al.} Phys.~Lett.~{\bf B205} 416 (1988).
\bibitem{feb98}Y.~Fukuda {\em et al.} Phys.~Lett.~{\bf B433} 9 (1998).
\bibitem{casper}D.~Casper {\em et al.} Phys.~Rev.~Lett.~{\bf 66} 2561 (1991).\\
R.~Becker-Szendy {\em et al.} Phys.~Rev.~{\bf D46} 3720 (1992).
\bibitem{kam92}K.~Hirata {\em et al.} Phys.~Lett.~{\bf B280} 146 (1992).
\bibitem{takita}M.~Takita, Ph.D. thesis, University of Tokyo (1989).
\bibitem{upward}Y.~Oyama {\em et al.} Phys.~Rev.~{\bf D39} 1481 (1989).
\bibitem{svob}R.~Becker-Szendy {\em et al.}  Phys.~Rev.~Lett.~{\bf 69}
1010 (1992).
\bibitem{frejus}Ch.~Berger {\em et al.} Phys.~Lett.~{\bf B245} 305 (1990).
\bibitem{nuosc}Y.~Fukuda {\em et al.} Phys.~Rev.~Lett.~{\bf 81} 1562 (1998).
\bibitem{dpf}M.~Messier, Talk presented at the 1999 DPF meeting,
January 1999.
http://hep.bu.edu/$\sim$messier/dpf/index.html\\
K.~Scholberg, hep-ex/9905016
\bibitem{vary}J.~LoSecco,  hep-ph/9903310
\bibitem{kasuga}S.~Kasuga Ph.D.~thesis, University of Tokyo (1998).
\end{thebibliography}
\end{document}